\documentclass[aps,prl,twocolumn,showpacs,letterpaper,superscriptaddress]{revtex4}
\usepackage{graphicx}
\usepackage{amsmath}
\usepackage{grffile}
\usepackage{color}
\def\beqn{\begin{eqnarray}}
\def\eeqn{\end{eqnarray}}
\def\beqns{\begin{eqnarray*}}
\def\eeqns{\end{eqnarray*}}
\def\beq{\begin{equation}}
\def\eeq{\end{equation}}
\def\bea{\begin{array}}
\def\ea{\end{array}}

\def\<{\langle}
\def\>{\rangle}

\newlength{\textlarg}

\begin{document}
\title{Conductivity of  graphene with resonant and non resonant adsorbates}
\author{Guy \surname{Trambly de Laissardi\`ere}}
\affiliation{Laboratoire de Physique th\'eorique et Mod\'elisation, CNRS and 
Universit\'e de Cergy-Pontoise, 
95302 Cergy-Pontoise, France}
\author{Didier \surname{Mayou}}
\affiliation{
Univ. Grenoble Alpes, Inst NEEL, F-38042 Grenoble, France \\
CNRS, Inst NEEL, F-38042 Grenoble, France
}
\date{\today}

\begin{abstract}

We propose a unified description of transport in  graphene with adsorbates that fully takes into account localization effects and loss of electronic coherence due to inelastic processes. We focus in particular on the role of the scattering properties of the adsorbates  and  analyze in detail cases  with resonant or non resonant scattering. For both models we identify  several regimes of conduction depending on the value of the Fermi energy. Sufficiently far from the Dirac energy and at sufficiently small concentrations the semi-classical theory can be  a good approximation. Near the Dirac energy we identify different quantum regimes,  where  the conductivity  presents  universal behaviors.

\end{abstract}

\maketitle



Electronic transport in graphene \cite{Berger04,Novoselov05,Zhang05,Berger06} is sensitive to static defects that are for example frozen ripples,  screened charged impurities,  or local defects like vacancies or adsorbates \cite{Hashimoto04,Wu07,Wu08,Zhou08}. Adsorbates, which can be organic groups or adatoms attached to the surface of graphene,  are of particular interest  in the context of functionalisation  which aims at  controlling the electronic properties by attaching atoms or molecules to  graphene \cite{Grassi,Peres06,Bostwick09,Fuhrer,Leconte10,Roche12}.  Therefore there is a need for a theory of conductivity in the presence of such defects. 

Theoretical studies of transport in the presence of  local defects have dealt mainly  either with the Bloch-Boltzmann formalism or with self-consistent approximations
\cite{Peres06,ANDO,Guinea08,Pereira08a,Robinson08,Wehling10,Skrypnyk10,Skrypnyk11,Ferreira11}. 
In these theories a major length scale that characterizes the electron scattering is the elastic mean-free path $L_{e}$. These approaches indeed explain some experimental observations such as  the quasilinear variation of conductivity with concentration of charge carriers  \cite{Peres06,Bostwick09,Fuhrer,Leconte10,Roche12}.  Yet these theories have important limitations and can hardly describe in detail the localization phenomena that has been reported in some experiments \cite{Wu07,Wu08,Bostwick09,Fuhrer}. 
Indeed in  the presence of a short range potential, such as that produced by  local defects  the electronic  states  are  localized on a length scale $\xi$ \cite{Ostrovsky10,Bang,Peres10,Roche13}. A sample will be insulating  unless some  source of scattering, like electron-electron  or electron-phonon interaction,  leads to a loss of the phase coherence on a length scale $L_{i} <\xi$. Therefore, in addition to the elastic mean-free path $L_{e}$,  the inelastic mean-free path $L_{i}$ and the localization length $\xi$  play also a fundamental role for the conductivity of graphene with adsorbates.


In this letter we develop a numerical approach for the  conductivity   that treats exactly the tight-binding Hamiltonian and  takes fully into account the effect of Anderson localization. This approach gives access to the characteristic lengths and to  the conductivity as a function of the concentration, the  Fermi energy $E_{\rm F}$ and   the inelastic mean-free path  $L_{i}$. In real samples $L_{i}$ depends on the temperature, or magnetic field, but it is an adjustable  parameter in this work. Our results confirm that sufficiently far from the Dirac energy, and for sufficiently small adsorbates  concentrations,  the Bloch-Boltzmann theory and  the self-consistent theories are  valid when  $L_{e} \ll  L_{i} \ll \xi$.  Near the Dirac energy we identify different regimes of transport that depend on whether the adsorbates produce resonant or non resonant scattering. These different regimes of transport present some universal characteristics which consequences are discussed for experimental measurements of conductivity and magneto-conductivity.

\textit{Models of adsorbates}

The scattering properties of local defects like adsorbates or vacancies is characterized by their T-matrix. Local defects tend to scatter electrons in an isotropic way for each valley  and  lead also to strong inter valley scattering. Yet the energy dependence of the T-matrix  depends very much on the type of defect and in this work we focus on the role of this energy dependence. To this end we consider two models for which  the T-matrix diverges at the Dirac energy (resonant adsorbates leading to mid-gap states also called zero energy modes)  or is constant (non resonant adsorbates). Note that resonances can occur also at non zero energy but here we restrict to the important case of zero energy modes. The conclusions drawn here, concerning the influence of the energy dependence of the T-matrix  for adsorbates,  are useful for other types of local defects. 

We consider that the adsorbates create a covalent bond  with some atoms of the graphene sheet.  Then a generic model  is obtained by removing the $p_z$ orbitals of these carbon atoms \cite{Pereira06,Pereira08a,Ostrovsky10,Bang,Peres10,Roche13,Robinson08,Wehling10,Ferreira11,Incze03,Lherbier12}. For example an hydrogen adsorbate  can be modeled by removing the $p_z$ orbital of the carbon atom that is just below the hydrogen atom. This is the model of resonant adsorbate that we consider here.  In this case the T-matrix associated to the adsorbate, diverges at the Dirac energy hence the name of resonant scatterers.  The non resonant model is constituted by two neighboring missing orbitals (divacancy). In that case the T-matrix is nearly constant  close to the Dirac energy and does not diverge. 

Finally we consider  here that the up and down spin are degenerate i.e. we deal with  a paramagnetic state. Indeed the existence of a magnetic state for various adsorbates, like hydrogen for example,  is still debated \cite{Nair12}.  Let us emphasize that  in the case of a magnetic state  the up and down spin  give two different contributions to the conductivity but the individual contribution of each spin can be analyzed from the results discussed here.  With these assumptions the generic model  Hamiltonian for adsorbates  writes:
\begin{equation}
H = - t \sum_{\<i,j\>}(c^{+}_{i}c_{j} +c^{+}_{j}c_{i})
\label{Ham}
\end{equation}
where $\<i,j\>$ represents nearest neighbours pairs of occupied sites and  $t = 2.7$\,eV determines the  energy scale. In our calculations the vacant sites (resonant adsorbates) or the di-vacant sites (non resonant adsorbates) are distributed at random with a finite concentration.

\textit{Evaluation of the  conductivity}

The present study relies upon  the Einstein relation between the conductivity and the quantum diffusion. We evaluate numerically the quantum diffusion using the MKRT approach \cite{Mayou88,Mayou95,Roche97,Roche99,Triozon02}.
This method has been used to study quantum transport  in disordered graphene, chemically doped graphene, graphene with functionalization
and graphene with structural defects \cite{Lherbier08,Lherbier08b,Leconte10,Trambly10,Leconte11,Lherbier11,Leconte11b,Lherbier12,Roche12,Roche13}. 
We introduce an inelastic scattering time $\tau_{i}$, beyond which the propagation becomes diffusive due to the destruction of coherence by inelastic processes (relaxation time approximation) 
\cite{Trambly06,Berger93,Belin93,Trambly05,Mayou00,Ciuchi11}. 
We finally get (Supplemental Material Sec. I):
\begin{eqnarray}
\sigma(E_{\rm F},\tau_{i})&=& e^{2} n(E_{\rm F})D(E_{\rm F},\tau_{i})\\
D(E_{\rm F},\tau_{i})&=&\frac{L_{i}^{2}(E_{\rm F},\tau_{i})}{2 \tau_{i}}
\label{eqf5}
\end{eqnarray}
where $E_{\rm F}$ is the Fermi energy, $n(E_{\rm F})$ the density of states (DOS),
$D(E_{\rm F},\tau_{i})$ the diffusivity, $\tau_{i}$ the inelastic scattering time and $L_{i}(E_{\rm F},\tau_{i})$  the inelastic mean-free path. $L_{i}(E_{\rm F},\tau_{i})$ is the typical distance of propagation  during the time interval $\tau_{i}$ for electrons at the energy $E_{\rm F}$ in the system without inelastic scattering  \cite{Lee85}.

The typical variation of $\sigma(\tau_{i})$ in our study (Supplemental Material Sec. II) is equivalent to that  found in previous works \cite{Roche12,Trambly10}. At small times the propagation  is ballistic and  the conductivity $\sigma(\tau_{i})$ increases when $\tau_{i}$ increases. For  large $\tau_{i}$   the conductivity $\sigma(\tau_{i})$ decreases with increasing $\tau_{i}$  due to quantum interferences effects and ultimately goes to zero in our case due to Anderson localization in 2 dimension.

We define the  microscopic conductivity $\sigma_{M}$ as the maximum value  of the conductivity over  all values of $\tau_{i}$. According to the renormalization theory this value is obtained when the inelastic mean-free path $L_{i}(\tau_{i})$ and the elastic mean-free path $L_{e}$ are comparable.  This microscopic conductivity $\sigma_{M}$ represents the conductivity without the effect of quantum interferences in the diffusive regime (localization effects)  and can be compared to semi-classical or self-consistent theories which also do not take into account the effect of quantum interferences  in the diffusive regime. 


Finally we note that in the above formulas it is assumed that the inelastic scattering does not affect the DOS $n(E_{\rm F})$. However this scattering can lead also to a mixing of states that affects the DOS. In the Supplemental Material (Sec. V) we analyze in detail this effect of mixing of states. Although it is difficult to quantify our results strongly suggest that the effect of the mixing plays a minor role except  for the microscopic conductivity $\sigma_{M}$ of the zero energy modes where indeed the DOS varies quickly. This leads us to conclusions  in contrast with those of a recent study  \cite{Roche13} (see below and Supplemental Material Sec. V).

\textit{Resonant adsorbates (monovacancies)}

\begin{figure}
\includegraphics [width=0.37\textwidth] {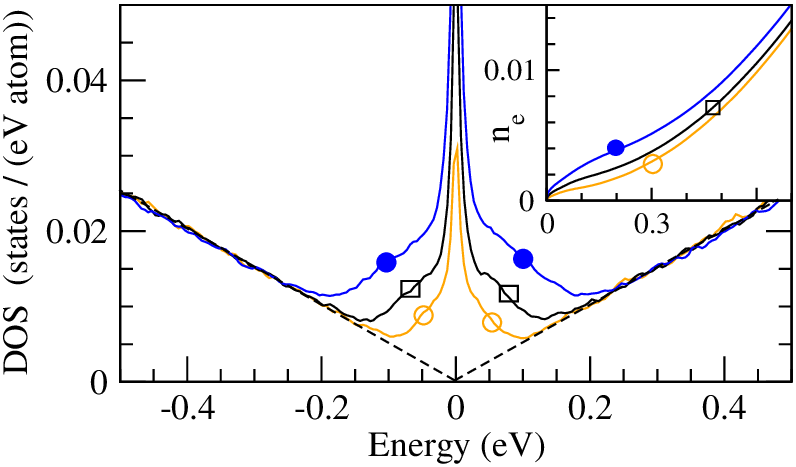}
\caption{Densities of states versus energy for resonant adsorbates (monovacancies) with concentrations (empty circles) $0.1\%$,  (empty square) $0.2\%$,  (filled circle) $0.4\%$. 
(Dashed lines) without adsorbate.
[Inset: electron density per atom $n_{e}$ versus energy.]
}
\label{f1b}
\end{figure}

Figure \ref{f1b} shows the total DOS  with three different regimes consistent with previous studies \cite{Peres06,Roche13}. At sufficiently large energies the density of pure graphene is weakly affected. Near the Dirac energy there is an intermediate regime where the pseudo-gap is filled. Very close to the Dirac point there is a third regime where the  density  presents a peak which is reminiscent of the mid-gap state (also called  zero energy modes)  produced by just one missing orbital.

\begin{figure}
~\includegraphics[clip,width=0.37\textwidth]{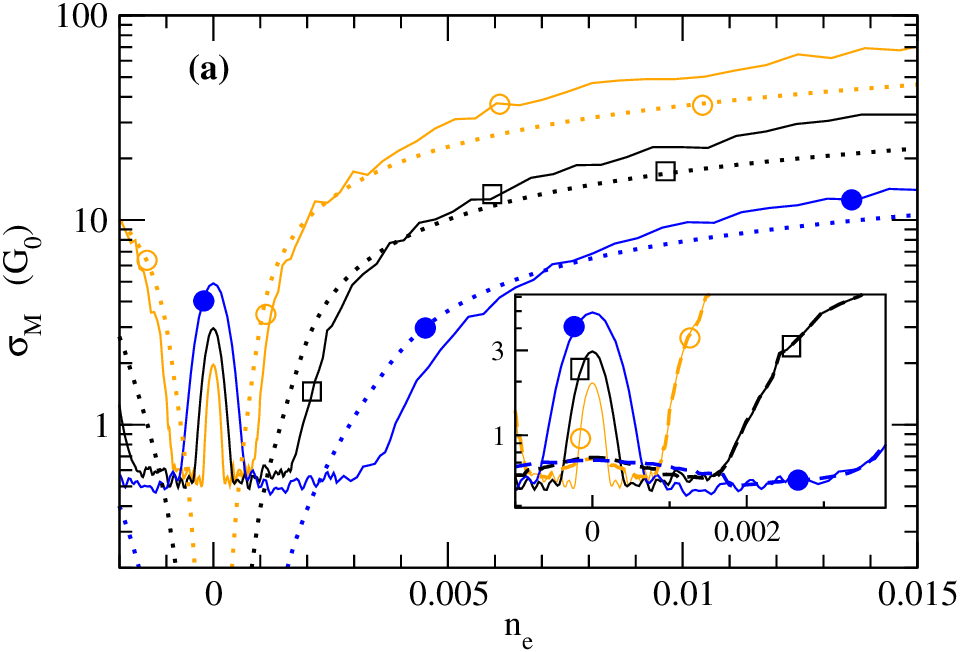}

\includegraphics[clip,width=0.37\textwidth]{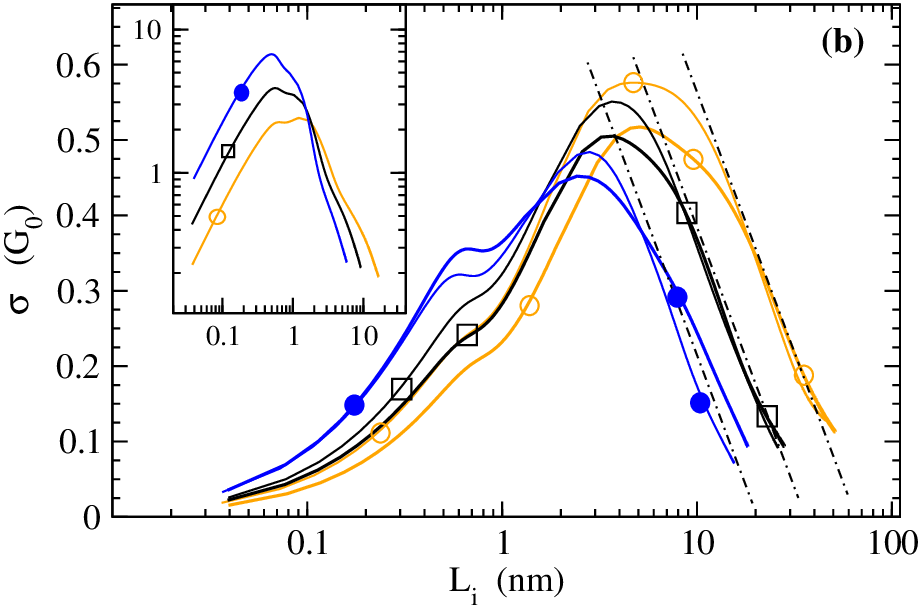}
\caption{Conductivity for 
resonant adsorbates (monovacancies) 
for 3 concentrations (see caption of figure \ref{f1b}). 
(a) Microscopic conductivity $\sigma_{M}$ versus electron density per atom $n_{e}$.  
(Dotted lines) predictions of the Boltzmann theory close to the Dirac energy  \cite{Wehling10}.
[Inset: zoom on the low concentration limit:
(lines) without mixing of states,
(dashed lines) with mixing of states on an energy range $\delta E = \hbar / \tau_i$ 
(Supplemental Material Sec. V).]
(b) Conductivity $\sigma$ versus inelastic scattering length $L_{i}$
at energies  $E=0.03$\,eV (thin line), and  $E=0.04$\,eV (thick line).  
The dot-dashed straight  lines show the slope $\alpha=0.25$  for $L_{i} \gg L_{e}$ (see text): 
[Inset: $\sigma(L_{i})$ at $E=0$ in a Log-Log scale.] 
$G_{0} = 2e^2/h$.
}
\label{f3b}
\end{figure}


Figure \ref{f3b} shows these three regimes for the microscopic conductivity $\sigma_{M}$. In the first regime i.e. at  sufficiently large energies,   $\sigma_{M} \simeq \sigma_{\rm B}$, where $\sigma_{\rm B}$ is calculated with the Bloch-Boltzmann approach \cite{Wehling10}. In this regime where the DOS is weakly affected (see above), 
$\sigma_{M} \gg G_{0} = 2e^2/h $ and $\xi \gg L_{e}$ (Supplemental Material Sec. III and IV). 

When the energy decreases, the semi-classical model fails (figure \ref{f3b}), and a second regime occurs  in which  $\sigma_{M}\simeq 4 e^{2} /\pi h$. This is consistent with predictions of self-consistent theories and with numerical calculations \cite{ANDO,Peres06,Guinea08,Pereira08a,Robinson08,Wehling10,Ferreira11,Roche13}.   
In agreement with the literature we find that the onset  for this regime corresponds to about one electron per impurity as shown by  figure \ref{f3b}. Also this intermediate regime occurs when the Fermi  wave-vector $k_{\rm F}$ 
is such that  $k_{\rm F}L_{e} \simeq 1$.

A characteristic length scale  in this intermediate regime is the distance $d$  between adsorbates (Supplemental Material). Here $d=1/\sqrt{n}$ where $n$ is the adsorbates density and $d\simeq 5$\,nm for a concentration of $0.1\%$.  In  this intermediate regime $L_{e}$ (defined precisely in the Supplemental Material Sec. III) depends on the energy but stays comparable to $d$. This can be understood by noting that $L_{e}$, which according to the semi-classical theory tends to zero at the  Dirac energy, cannot be much smaller than the distance $d$  between the scattering centers.

At smaller energies a peak of the microscopic conductivity $\sigma_{M}$ appears very close to the Dirac energy which coincides with the peak of the DOS  and represents a third regime of transport.  This peak of conductivity is not predicted by self-consistent theories. It is not obtained by \cite{Robinson08,Wehling10} and is present in the calculation of \cite{Ferreira11} although much less marked than in the present work. This peak is obtained with very similar values in the recent work  \cite{Roche13}. In this peak $\sigma_{M}$ increases with the concentration of defects. Yet $\sigma_{M}$ is calculated here by neglecting the mixing of energy levels due to the inelastic scattering processes.  As shown in the inset of figure \ref{f3b}a we find that this peak can decrease when  the mixing of the levels due to the inelastic scattering processes is taken into account (Supplemental Material Sec. V).





We discuss  now these three regimes for the conductivity when  $ L_{i} $  $>$  $ L_{e}$ (figure \ref{f3b}b). At high energies we find standard localization effects consistent with very large localization lengths (Supplemental Material). In the intermediate regime  (i.e. $\sigma_{M} \simeq 4 e^{2} /\pi h$), for concentration $0.1\%$ to $10\%$,   
the conductivity is well represented by  the equation:

\beq
\sigma (L_{i})\simeq \frac{4 e^{2}}{\pi h} -\alpha \frac{2e^{2}}{h} {\rm Log}\left( \frac{L_{i}}{L_{e}} \right).
\label{sigma}
\eeq

The coefficient  is $\alpha \simeq 0.25$ which is close to the result of the perturbation theory of 2 dimension Anderson localization for which  $\alpha \simeq 1/\pi$  \cite{Lee85}. We emphasize that the regime is not perturbative close to the Dirac point. 
This expression shows no effect of anti localization \cite{McCann06} as expected for purely short range scattering. Indeed in that case graphene belongs to an orthogonal symmetry class with localization effects as in standard 2 dimension metal without spin-orbit coupling \cite{Mucciolo10}. 
The localization length $\xi$ deduced from this expression (\ref{sigma}) is such that $\sigma (\xi) =0$ which gives $\xi \simeq 13 L_{e}$.  
This results justifies previous estimates  of the localization length from the calculation of the elastic mean-free path that were done in this plateau of microcopic conductivity \cite{Lherbier11}.
As discussed above the elastic mean-free path $L_{e}$ depends on the energy in this regime but  is of the order of the distance $d$ between adsorbates. Therefore $d$  determines the order of magnitude of the localization length $\xi$ and of the elastic mean-free path $L_{e}$. More precisely in the range of concentration $0.1\%$ to $10\%$ 
(Supplemental Material Sec. IV) 
the localization length in this regime is $\xi \simeq 20 d/r$  where the ratio $r$ decreases with increasing energy and is  $1\leq r \leq 3$. The values of the localization length found at $10\%$ in \cite{Bang} are consistent with our study.

In the third regime where the DOS  and $\sigma_{M}$  present a peak, the conductivity does not follow the above law (equation (\ref{sigma})). $\sigma (L_{i} )$ fits better with a power law $\sigma (L_{i} ) \propto  L_{i}^{-\beta} $ where $\beta$ depends on the concentration (here $1< \beta <2$). This is consistent with the divergence of the localization length $\xi$ predicted in  \cite{Ostrovsky10} although we do not recover the behavior found precisely at the Dirac energy.  Since our energy resolution is of the order of $10^{-2}$\,eV, we conclude that the zero energy behavior of the conductance  exists only in a narrow energy range and could be difficult to observe experimentally.

In the presence of a magnetic field the magnetic length $L(B)=\sqrt{\hbar/eB}$ plays the role of a finite coherence length just  as the inelastic mean-free path $ L_{i} (T)$.  When  $L(B)<L_{i} (T)$  the relevant coherence length is $L(B)$ and the conductivity is  $\sigma(L(B))$. This could be compared to our results, in particular  for equation (\ref{sigma}).

\textit{Non resonant adsorbates (divacancies)}

\begin{figure}
\includegraphics [width=0.37\textwidth] {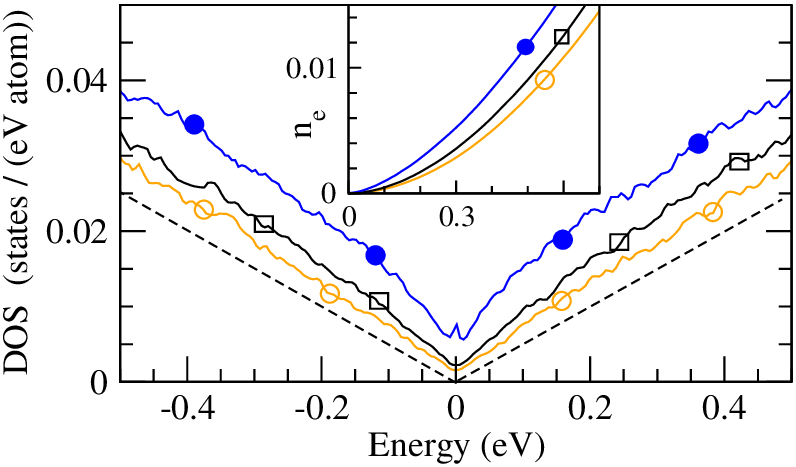}
\caption{Densities of states versus energy for non resonant adsorbates (divacancies) with concentrations $0.5\%$ (empty circles), $1\%$ (empty squares),  $2\%$ (filled circle). 
(Dashed lines) without adsorbate.
[Inset: electron density per atom $n_{e}$ versus energy.]
}
\label{f1a}
\end{figure}

\begin{figure}
\includegraphics[clip,width=0.37\textwidth]{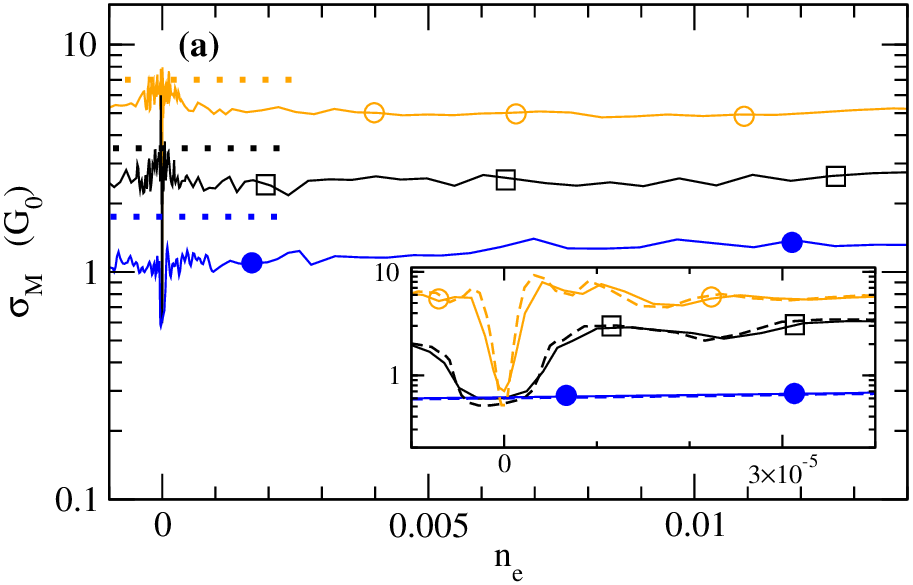}

\includegraphics[clip,width=0.37\textwidth]{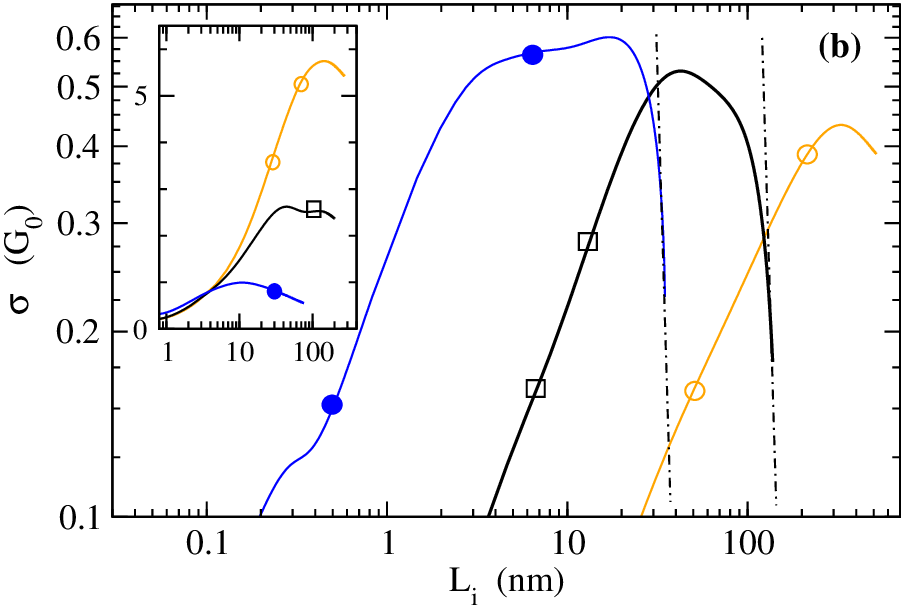}
\caption{Conductivity for 
 non resonant adsorbates (divacancies) 
for 3 concentrations (see caption of figure \ref{f1a}).  
(a) Microscopic conductivity $\sigma_{M}$ versus electron density per atom $n_{e}$.
(Dotted lines) predictions of the Boltzmann theory close to the Dirac energy.
[Inset: zoom on the low concentration limit:
(lines) without mixing of states,
(dashed lines) with mixing of states on an energy range $\delta E = \hbar / \tau_i$ 
(Supplemental Material Sec. V).]
(b) Conductivity $\sigma$  versus inelastic scattering length $L_{i}$
at  $E=0$ in a Log-Log scale. 
[Inset:  $\sigma(L_{i})$ at $E=0.1$\,eV.] $G_{0} = 2e^2/h$.
}
\label{f3a}
\end{figure}


Figure \ref{f1a} shows the total densities of states  as a function of energy for the  non resonant adsorbates. The result  is similar to that obtained by the self-consistent Born approximation (SCBA)  for Anderson disorder \cite{ANDO}. The two models are not strictly equivalent but have both an energy independent T-matrix close to the Dirac energy.  The microscopic conductivity  $\sigma_{M}$ presents a minimum with $\sigma_{M}\simeq 4 e^{2} /\pi h$  in a narrow concentration range (figure \ref{f3a}).  Again this is consistent with the SCBA predictions for the Anderson model \cite{ANDO}.  At the Dirac energy we find that the conductivity can be represented by a power law  $\sigma (L_{i} ) \propto  L_{i}^{-\gamma} $ with $\gamma \simeq 4-6$. Yet the equation (\ref{sigma}) fits also with $\alpha \simeq 0.75$ which gives $\xi\simeq 2.5 L_{e}$. In any case the quick decrease of the conductivity $\sigma (L_{i} )$ with $L_{i} $ and the narrow concentration range for the minimum of $\sigma_{M}$ suggest that the value $\sigma_{M}\simeq 4 e^{2} /\pi h$ could be very difficult to find experimentally. 

A recent experimental work \cite{Nakaharai13} shows that  graphene with defects induced by helium ion, at about  1\% concentration,  presents Anderson localization even at room temperature. Our study suggest that at such concentration only resonant adsorbates can create the strong localization. The length of the samples, less than 100\,nm, is also consistent with small inelastic scattering \cite{Wu07}.


\textit{Conclusion}

To conclude our study  shows that  the energy dependence of the scattering properties  of local defects is determinant for transport and magneto transport properties of graphene with adsorbates.  Sufficiently far from the Dirac energy, and for not too high concentrations, the semi-classical approach is  usually valid. Yet closer from the Dirac point there are regimes where the quantum effects are essential. For resonant adsorbates we find that in the regime of the so-called minimum conductivity the conductivity is well represented by equation (\ref{sigma}). The characteristic length scale is  the distance $d$ between defects and the localization length $\xi$ and the elastic mean-free path $L_{e}$ are given by $\xi\simeq 13 L_{e} \simeq 20 d/r$  where the ratio $r$ decreases with increasing energy and is  $1\leq  r  \leq 3$. Closer from the Dirac energy there is a peak in the DOS which corresponds to another  regime of transport in a band of mid-gap states.  We find a  critical behaviour partly consistent with \cite{Ostrovsky10,Roche13}. In this regime we have shown that the inelastic scattering can destroy the peak of DOS, which strongly affects the conductivity. Yet a proper understanding of the physics of transport  in this peak requires clearly  further studies \cite{Ducastelle13}. For non resonant adsorbates, in a narrow energy range near the Dirac energy the microscopic conductivity $\sigma_{M}$ presents a minimum with the universal value  $\sigma_{M} \simeq 4 e^{2}/\pi h$. Yet at the Dirac energy  there are   strong localization effects. These could make the experimental observation of the the universal value  $\sigma_{M} \simeq 4 e^{2}/\pi h$ very difficult. Finally we emphasize that the methodology used here to study quantum effects for electronic  transport is of wide applicability. In particular important related problems such as magneto-conductivity of graphene beyond low field  limit or competition between scattering by defects with long range and short range potential  could be studied as well \cite{McCann06}.



\textit{Acknowledgements}

We thank  L. Magaud, C. Berger and W. A. de Heer for fruitfull discussions and comments.
The computations have been performed at the
Centre de Calcul of the 
Universit\'e de Cergy-Pontoise.
We thank Y. Costes and D. Domergue for computing assistance.


\vskip 1cm
\begin{center}
{\Large \bf
Supplemental Material
}
\end{center}

\section{I. Evaluation of the Kubo-Greenwood conductivity}

The present study relies upon the evaluation of the Kubo-Greenwood conductivity using the Einstein relation between the conductivity and the quantum diffusion
(see Refs. \cite{Mayou00,MayouRevueTransp} and Refs. therein).
A central quantities are the velocity correlation function of
states of energy $E$ at time $t$,
\begin{eqnarray}
C(E,t) &=& \Big\langle \hat{V}_x(t)\hat{V}_x(0) + \hat{V}_x(0)\hat{V}_x(t) \Big\rangle_E \\
&=& 2\,{\rm Re}\, \Big\langle \hat{V}_x(t)\hat{V}_x(0) \Big\rangle_E \, ,
\label{EqAutocorVit}
\end{eqnarray}
and the average
square spreading (quantum diffusion) of states of energy $E$ at time $t$
along the $x$ direction,
\begin{equation}
X^{2}(E,t)= \left \langle (\hat{X}(t)- \hat{X}(0)^{2} \right \rangle_{E}.
\label{XET}
\end{equation}
In equations (\ref{EqAutocorVit}) and  (\ref{XET}), 
$\left \langle ... \right \rangle_{E}$ is the average on states with energy $E$,  
Re$\,{A}$ is the real part of ${A}$,
$\hat{V}_x(t)$ and $\hat{X}(t)$ are the Heisenberg representation of the velocity operator $\hat{V}_x$ 
and the position operator $\hat{X}$
along $x$ direction at time $t$,
\begin{eqnarray}
\hat{V}_x = \frac{1}{i \hbar}~ \Big[ \hat{X} , \hat{H} \Big].
\end{eqnarray}
$C(E,t)$ is related to quantum
diffusion by the relation \cite{Mayou00},
\begin{eqnarray}
\frac{\rm d}{{\rm d} t} \Big(X^2(E,t) \Big) = \int_0^{t}C(E,t'){\rm d} t'.
\label{EqX2}
\end{eqnarray}
From Kubo-Greenwood formula, 
the conductivity is given by the Einstein relation,
\begin{eqnarray}
\sigma(E_{\rm F}) = e^2 n(E_{\rm F}) D(E_{\rm F}),
\label{EinsteinRelation}
\end{eqnarray}
where $e$ is the electron charge, $E_{\rm F}$ the Fermi energy,  $n$ the density of states and $D$ the diffusivity related 
to the square spreading by the relation \cite{MayouRevueTransp},
\begin{eqnarray}
D(E_{\rm F}) = \frac{1}{2} \, \lim_{t\rightarrow \infty} \frac {{\rm d}}{{\rm d} t} \,  X^2(E_{\rm F},t).
\label{EqDiffusivity}
\end{eqnarray}

We evaluate numerically the quantum diffusion $X^{2}(E,t )$ of states of energy $E$ for the Hamiltonian  using the MKRT approach \cite{Mayou88,Mayou95,Roche97,Roche99,Triozon02}. 
This method 
allows very efficient numerical calculations by recursion in real-space.
Our calculations are performed on samples containing up to $10^{8}$ atoms which corresponds to a  typical size of about one micron square. This allows to study systems with characteristic  inelastic mean-free path $L_{i}$ of the order of a few hundreds nanometers. 
With characteristic lengths of such size it is possible to treat systems with low concentrations of adsorbates that are of  $0.1\%,  0.2\%,  0.4\%$ for resonant adsorbates (monovacancies) 
and of $0.5\%,  1\%,  2\%$ for non-resonant adsorbates (divacancies). 
For the results presented here the energy resolution is of the order of $10^{-2}$\,eV.

The effect of the inelastic scattering is treated in a phenomenological way. 
This relaxation time approximation (RTA) has been used succesfully to compute \cite{Trambly06} 
conductivity in approximants of Quasicrystal where quantum diffusion and localization effect play a essential role \cite{Berger93,Belin93,Trambly05} and conductivity in organic semiconductors \cite{Ciuchi11}.
Following previous works \cite{Mayou00,MayouRevueTransp,Trambly06,Ciuchi11}, we assume that the velocity correlation function $C_{i}(E,t)$ of the system with inelastic scattering  is given by,
\begin{eqnarray}
C_{i}(E,t) ~\simeq ~ C(E,t)\, {\rm e}^{-|t|/\tau_{i}} \, ,
\label{RTA_C}
\end{eqnarray}
where $C(E,t)$ is the velocity correlation of the system with elastic scattering (monovacancies or divacancies) 
but without inelastic scattering. 
Here the inelastic scattering time $\tau_i$ is the cutoff time of the weak localization effects also called dephasing time.
As shown in Refs.\,\cite{Mayou00,MayouRevueTransp,Trambly06,Ciuchi11}
the propagation given by this formalism is unaffected by inelastic scattering at short times ($t < \tau_{i}$) and diffusive at long times ($t> \tau_{i}$) as it must be.  
Using the $t=0$ conditions, $X^2(E,t=0)=0$ and $\frac{{\rm d}}{{\rm d}t}X^2(E,t=0)=0$, 
and performing two integrations by part,
we obtain from equations
 (\ref{EqX2}), (\ref{EinsteinRelation}), (\ref{EqDiffusivity}) and (\ref{RTA_C}), \cite{MayouRevueTransp}
\begin{eqnarray}
\sigma(E_{\rm F},\tau_{i})&=& e^{2} n(E_{\rm F})D(E_{\rm F},\tau_{i}) \label{EquationEinsteinSM} \, ,\\  
D(E_{\rm F},\tau_{i})&=&\frac{L_{i}^{2}(E_{\rm F},\tau_{i})}{2 \tau_{i}} \, ,\\
L_{i}^{2}(E_{\rm F},\tau_{i})&=&\frac{1}{\tau_{i}} \int_0^\infty \! X^{2}(E_{\rm F},t)\,{\rm e}^{-t/\tau_{i}} \, \mathrm{d}t \, ,
\label{EquationLi}
\end{eqnarray}
where 
$L_{i}(E_{\rm F},\tau_{i})$  the inelastic mean-free path
and $D(E_{\rm F},\tau_{i})$ the diffusivity. 
$X^{2}(E,t )$ is calculated for the system with the Hamiltonian given in the main text  which represents  only elastic scattering due to monovacancies or divacancies. The above equations treat the inelastic scattering in a way that is equivalent to the standard approximation in mesoscopic physics. Indeed, in the presence of inelastic scattering,  it is usually assumed that,
$L_{i}^{2}(E_{\rm F}) \simeq \sqrt{X^{2}(E_{\rm F},\tau_{i})}$, thus 
the conductivity is given by the Einstein formula with a diffusivity   
$D(E_{\rm F},\tau_{i}) \simeq X^{2}(E_{\rm F},\tau_{i})/(2 \tau_{i})$ \cite{Lee85},  
which is essentially equivalent to the above equations.

\section{II. Conductivity and coherence length versus inelastic mean-free time}

\begin{figure}
\includegraphics[width=0.45\textwidth]{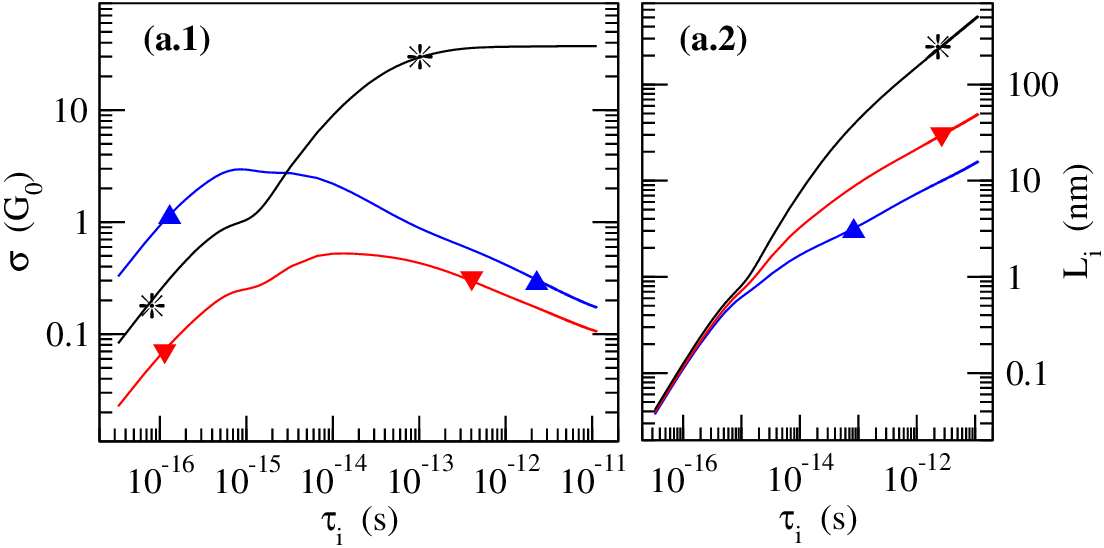}

\vskip .1cm
\includegraphics[width=0.45\textwidth]{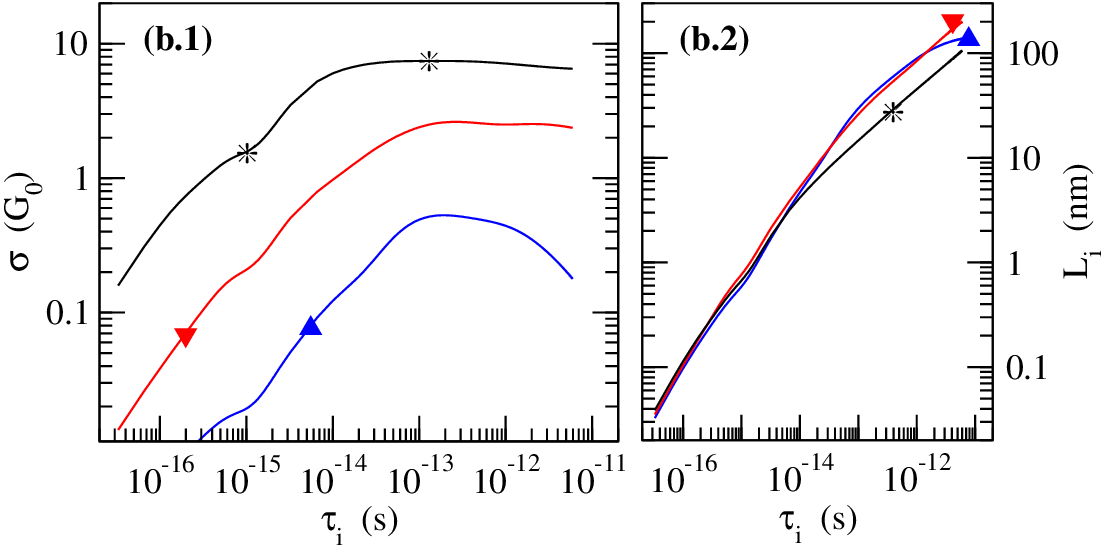}
\caption{ 
Conductivity $\sigma(E,\tau_{i})$ and inelastic mean-free path $L_{i}(E,\tau_{i})$
versus inelastic scattering time $\tau_{i}$.
(a) Concentration $0.2\%$ of resonant adsorbates (monovacancies) for energies
$E=0$ (triangle up),  $E=0.04$\,eV (triangle down) and  $E=0.8$\,eV (star).
(b) Concentration $1\%$ of non resonant adsorbates (divacancies) for energies 
$E=0$ (triangle up), $E=0.1$\,eV (triangle down) and  $E=1.5$\,eV (star).
$G_{0} = 2e^2/h$.
}
\label{f1}
\end{figure}

As explained in the main text we introduce an inelastic scattering time $\tau_{i}$, beyond which the propagation becomes diffusive due to the destruction of coherence by inelastic processes. Figure \ref{f1} presents the conductivity  $\sigma(E,\tau_{i})$ and the  inelastic mean-free path $L_{i}(E,\tau_{i})$ calculated as a function of inelastic scattering time $\tau_{i}$  for different energies $E$. 
Here $L_{i}(E,\tau_{i})$  is the length beyond which the propagation becomes diffusive due to the destruction of coherence by inelastic processes (equations (\ref{EquationEinsteinSM}) and (\ref{EquationLi})). Conductivities are calculated for large values of $\tau_{i}$ up to a few  $10^{-11}$\,s which corresponds to inelastic mean-free paths up to several hundred nanometers. This is  possible because we can treat large systems containing up to $10^{8}$ atoms. We have checked the convergence of our calculations with respect to the size of the sample. 

At small times the propagation  is ballistic and  the conductivity $\sigma(\tau_{i})$ increases when $\tau_{i}$ increases. For  large $\tau_{i}$   the conductivity $\sigma(\tau_{i})$ decreases with increasing $\tau_{i}$  due to quantum interferences effects and ultimately goes to zero in our case due to Anderson localization in 2 dimension.

We define the  microscopic conductivity $\sigma_{M}$ as the maximum value  of the conductivity over  all values of $\tau_{i}$. According to the renormalization theory this value is obtained when the inelastic mean-free path $L_{i}(\tau_{i})$ and the elastic mean-free path $L_{e}$ are comparable.  This microscopic conductivity $\sigma_{M}$ represents the conductivity without the effect of quantum interferences in the diffusive regime (localization effects). Therefore $\sigma_{M}$  be compared to semi-classical or self-consistent theories which also do not take into account the effect of quantum interferences  in the diffusive regime.

\begin{figure}[!]
\includegraphics[width=0.38\textwidth]{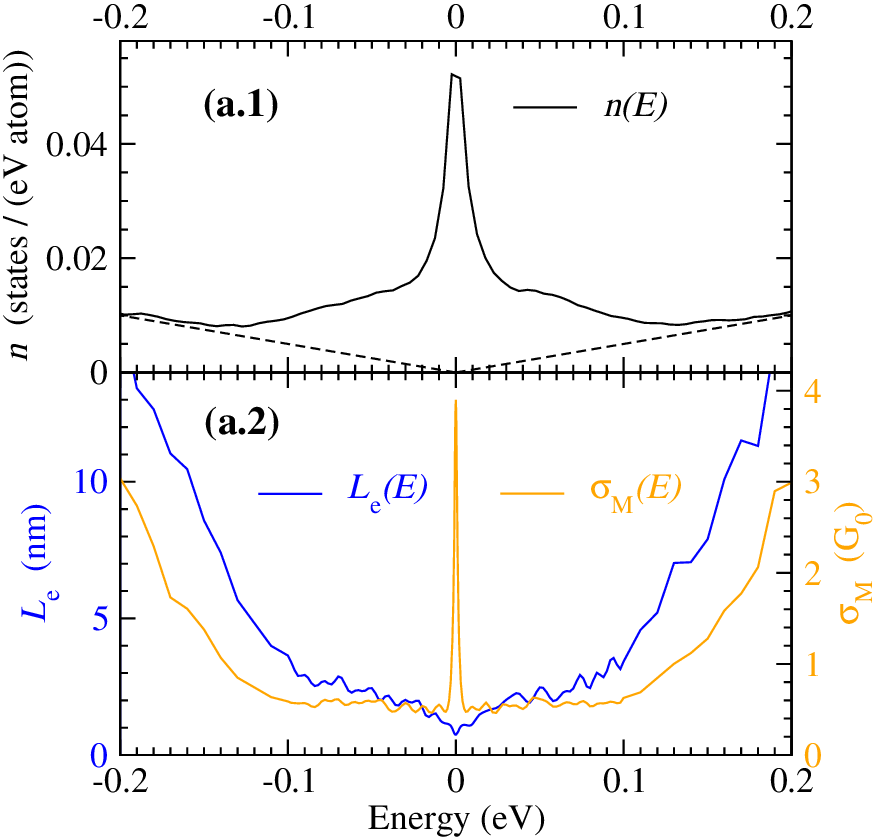}

\vskip .1cm
\includegraphics[width=0.38\textwidth]{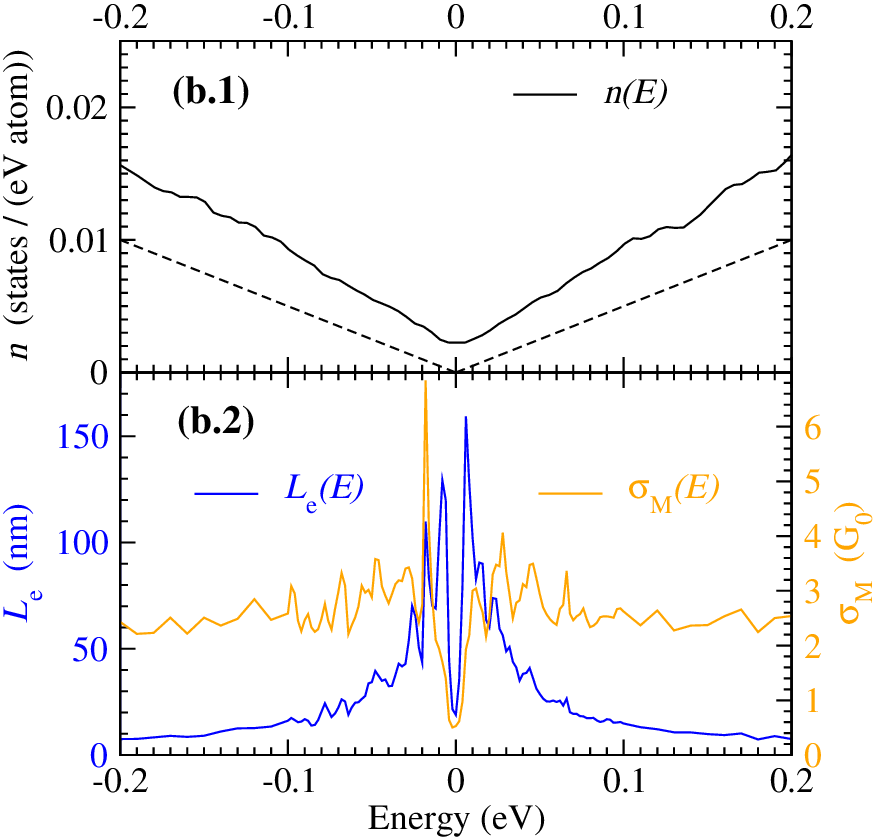}
\caption{ 
Density of states $n$, elastic mean-free path $L_e$ and 
microscopic conductivity $\sigma_M$ 
versus energy $E$.
(a) Concentration $0.2\%$ of resonant adsorbates (monovacancies).
(b) Concentration $1\%$ of non resonant adsorbates (divacancies).
(dashed lines) density of states of pure graphene.
The elastic mean-free path is computed from the maximum of the diffusivity $D_{M}(\tau_{\rm i})$ by
$L_e = 2D_{M} / V_{\rm F}$, with Fermi velocity $V_{\rm F} = 10^6$\,m.s$^{-1}$.
$G_{0} = 2e^2/h$.
}
\label{figsup1}
\end{figure}

\section{III. Elastic mean-free path}

We define the elastic mean-free path as $L_e = 2D_{M} / V_{\rm F}$ where $D_{M} $ is the microscopic diffusivity and $V_{\rm F}$ the Fermi velocity. In figure \ref{figsup1} we give two typical results for monovacancies and divacancies for the variation the elastic mean-free path with energy. 

For $0.2\%$ of resonant adsorbates (monovacancies)  the mean-free path decreases with energy up to $E\simeq 0.1$\,eV which is the onset of the plateau of the microscopic conductivity.  At $E= 0.1$\,eV,  $L_e$  is of the order of 4\,nm which is comparable to the distance between defects $d\simeq 3.5$\,nm. When the energy decreases the mean-free path decreases also but stays of the same order and  does not go to zero as predicted by the semi-classical theory. A qualitative explaination is that  the elastic mean-free path cannot be  much smaller than the distance between the scattering centers. This behavior is obtained at all the concentrations that we have studied. We find that, for energies in the intermediate regime (plateau of the microscopic conductivity), the elastic mean-free path is of the order of $L_e \simeq 3d/2 r$ where the ratio $r$ decreases with increasing energy and varies in the range  $r=1-3$.

For $1\%$ of non resonant adsorbates (divacancies) the microscopic conductivity is essentially constant at high energies in agreement with the semi-classical and self-consistent theories. In this regime the elastic mean-free path varies roughly like the inverse of the energy. Yet close to the Dirac energy the mean-free path abruptly decreases just as the microscopic conductivity. This is in agreement  with the self-consistent theory of Anderson disorder \cite{ANDO} that predicts an abrupt decrease of conductivity in a region where the density of states does not vary much. This means that  the microscopic diffusivity $D_{M}$ varies quickly and therefore the mean-free path $L_e = 2D_{M} / V_{\rm F}$ varies also quickly.

\section{IV. Intermediate regime for mono vacancies: effect of localization on the conductivity}

\begin{figure}[!]
\includegraphics[clip,width=0.4\textwidth]{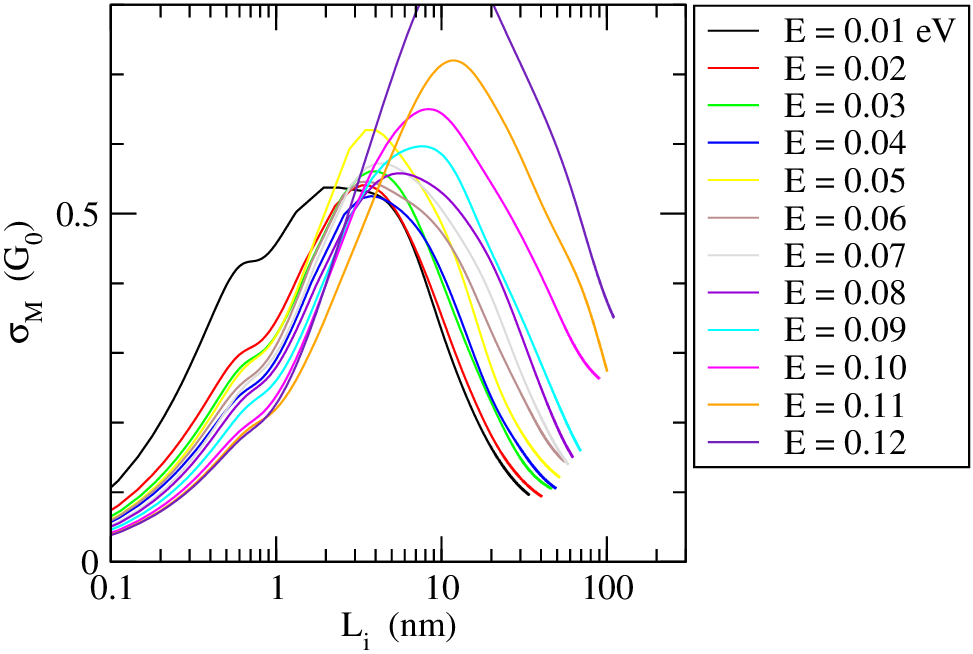}

\caption{Conductivity $\sigma$ in unit of  $G_{0} = 2e^2/h$ as a function of inelastic scattering length $L_{i}$
for 0.2\% of resonant adsorbates (monovacancies) at various energies  $E$.  
}
\label{figsup5}
\end{figure}

\begin{figure}[!]
\includegraphics[clip,width=0.35\textwidth]{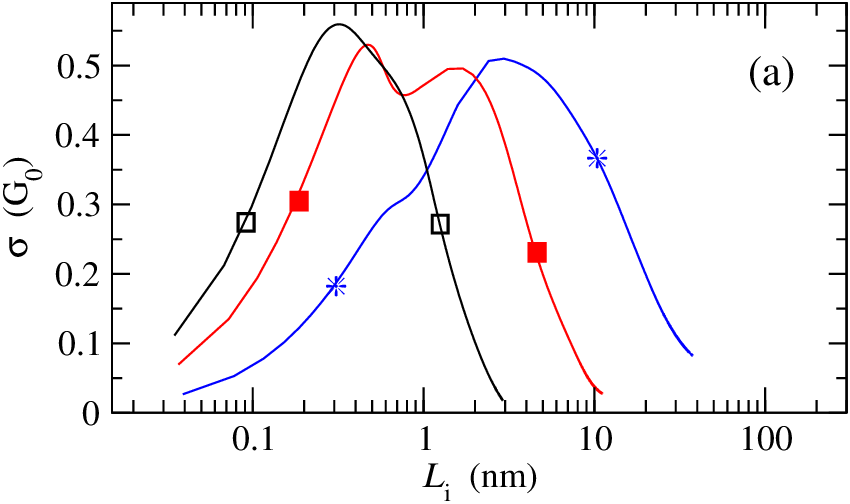}

\vskip .1cm
\includegraphics[clip,width=0.35\textwidth]{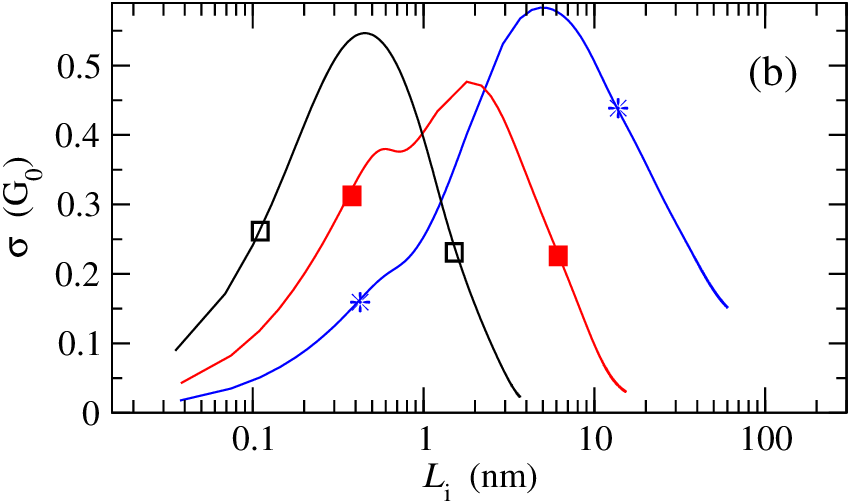}

\vskip .1cm
\includegraphics[clip,width=0.35\textwidth]{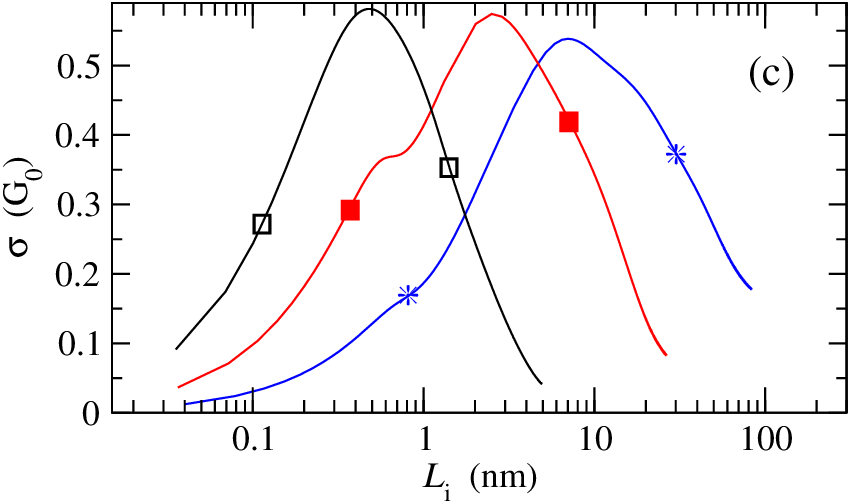}

\caption{Conductivity $\sigma$ in unit of  $G_{0} = 2e^2/h$ as a function of inelastic scattering length $L_{i}$
for (empty square) 0.1\%, (filled square) 1\%, (star) 10\% of resonant adsorbates (monovacancies) at energies:
(a) minimum energy of the intermediate regime, (b) middle energy the intermediate regime, (c) maximum energy of the intermediate regime.  
}
\label{figsup6}
\end{figure}

In the intermediate regime of monovacancies the microscopic conductivity  $\sigma_{M}$  is of the order of  $4 e^{2} /\pi h$.  We find a universal law for the conductivity as a function the inelastic mean-free path $ L_{i} $ and  the elastic mean-free path $ L_{e}$  with  $ L_{i} > L_{e}$ (see main text equation (\ref{sigma})):
\beq
\sigma (L_{i})\simeq \frac{4 e^{2}}{\pi h} -\alpha \frac{2e^{2}}{h} {\rm Log}\left( \frac{L_{i}}{L_{e}} \right) \, .
\label{sigmaSM}
\eeq
The coefficient  is $\alpha \simeq 0.25$.  Here we give additional results as compared to the main text. Figure \ref{figsup5} shows the results for $\sigma (L_{i})$  for 0.2\% of resonant adsorbates in a series of energies chosen in the region of the plateau (see figure \ref{figsup1}). One sees that the variation of conductivity is in agreement with equation (\ref{sigmaSM}) except for the lowest energy $E=0.01$\,eV and the highest $E>0.1$\,eV. Indeed these energies mark the lower bound and upper bound  of the intermediate regime.  We have checked that this law (equation (\ref{sigmaSM})) is valid even at higher concentration up to $10$\% of monovacancies (figure \ref{figsup6}).

\begin{figure}[!]
\includegraphics[width=.37\textwidth]{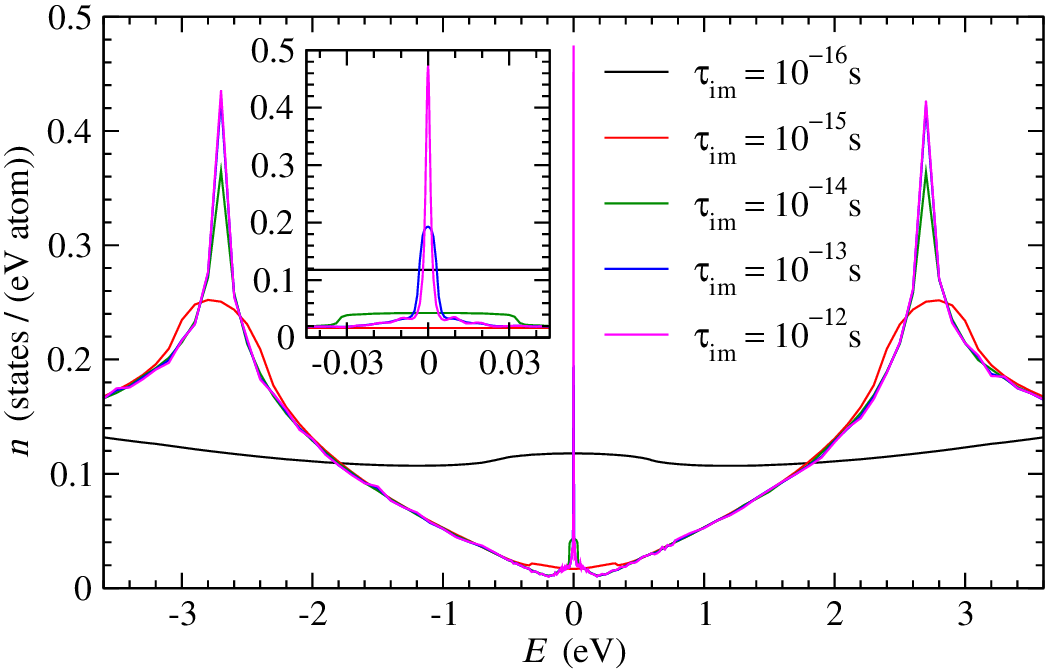}

\vskip .2cm
\includegraphics[width=.37\textwidth]{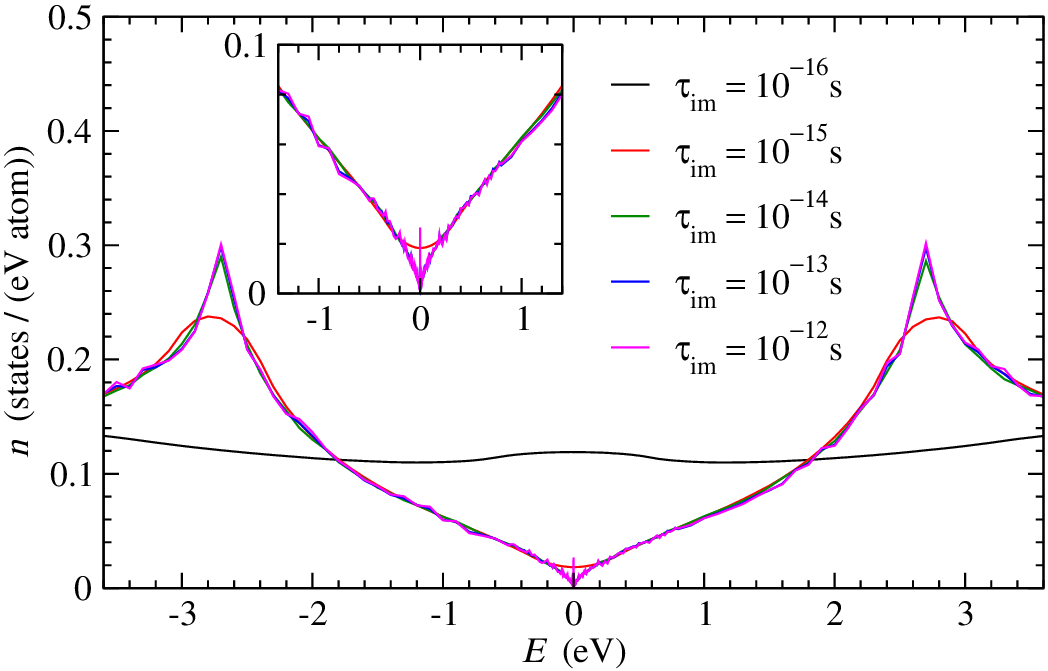}
\caption{ 
Density of states $\left \langle n(E) \right \rangle_{\eta} $ 
versus energy $E$ for various values of mixing energy $\delta E=\hbar/\tau_{im} = \eta \hbar/\tau_{i}$ (see text).
Upper panel: Concentration $0.4\%$ of resonant adsorbates (monovacancies).
Lower panel: Concentration $2\%$ of non resonant adsorbates (divacancies). At this concentration small clusters of adsorbates exist that create a peak in the density of states. This peak exists only for sufficiently long inelastic scattering time.
}
\label{figsup2}
\end{figure}

\begin{figure}[!]
\includegraphics[width=.45\textwidth]{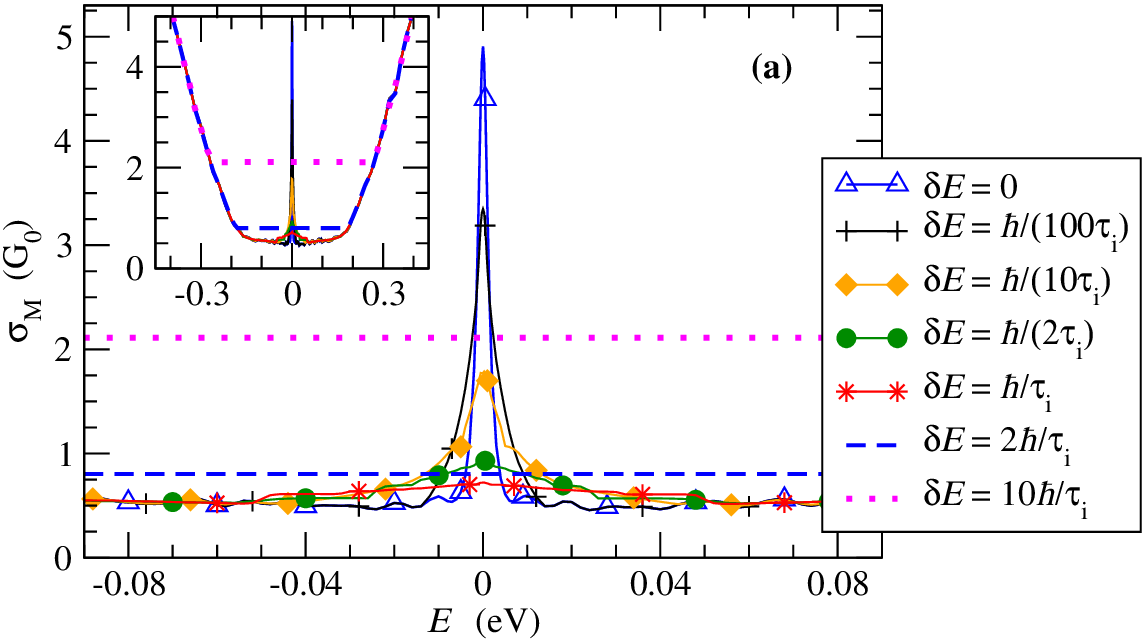}

\vskip .2cm
\includegraphics[width=.47\textwidth]{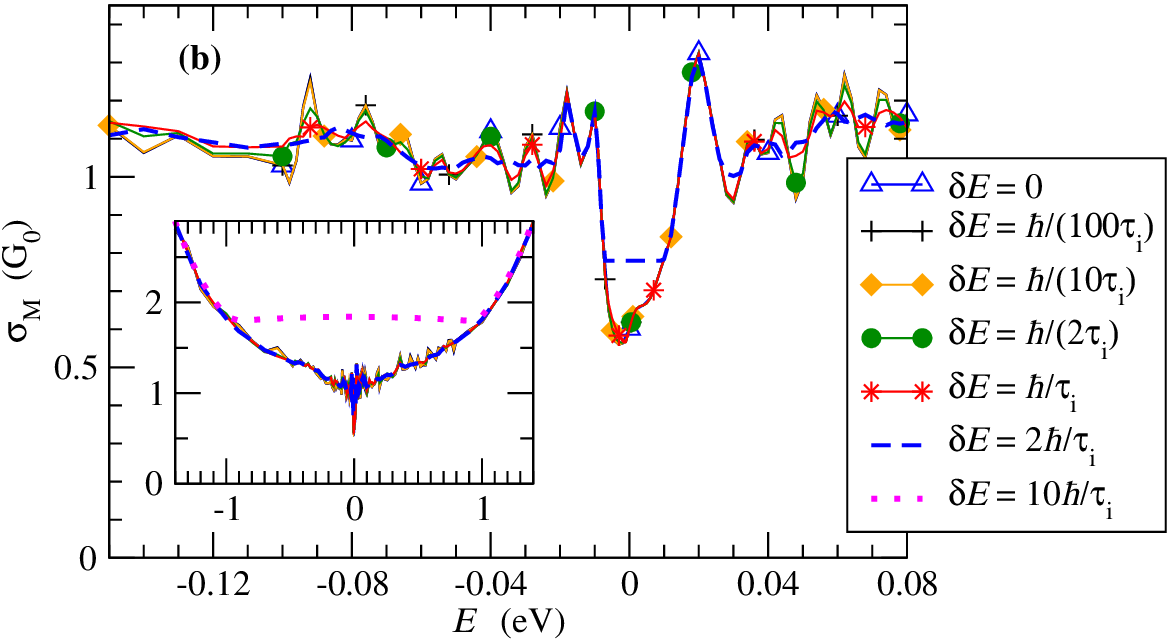}
\caption{ 
Microscopic conductivity $\left \langle  \sigma_M(E)  \right \rangle_{\eta} $ in unit of  $G_{0} = 2e^2/h$
versus energy $E$, for various values of the mixing energy $\delta E$, 
corresponding to $\eta$ varying from $1/100$ to 10 (see text).
(a) Concentration $0.4\%$ of Resonant adsorbates (monovacancies).
(b) Concentration $2\%$ of non resonant adsorbates (divacancies).
}
\label{figsup3}
\end{figure}

\section{V. Effect of the mixing of states induced by the inelastic scattering}

\begin{figure}
\includegraphics[width=0.47\textwidth]{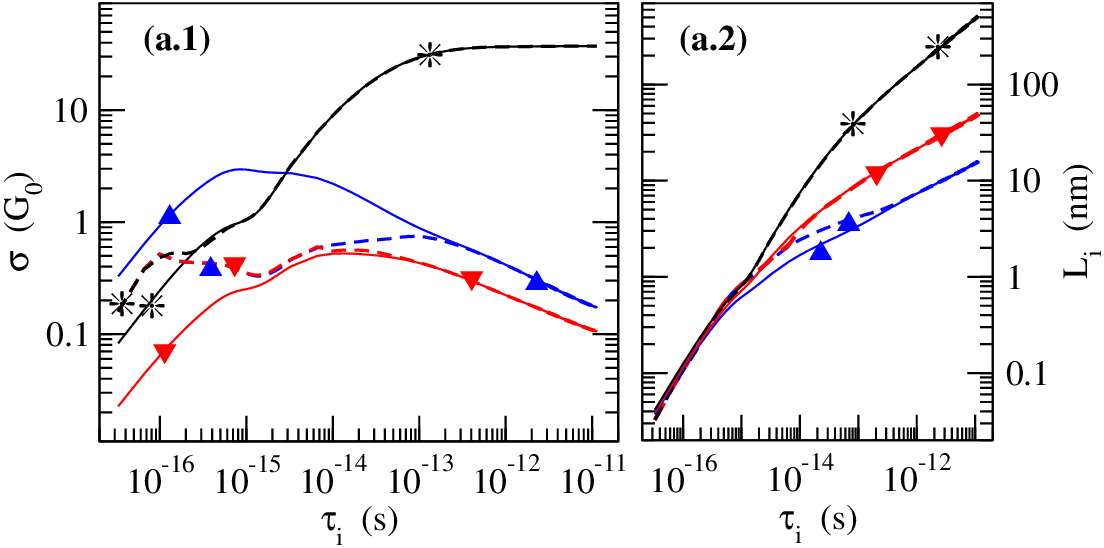}

\vskip .1cm
\includegraphics[width=0.47\textwidth]{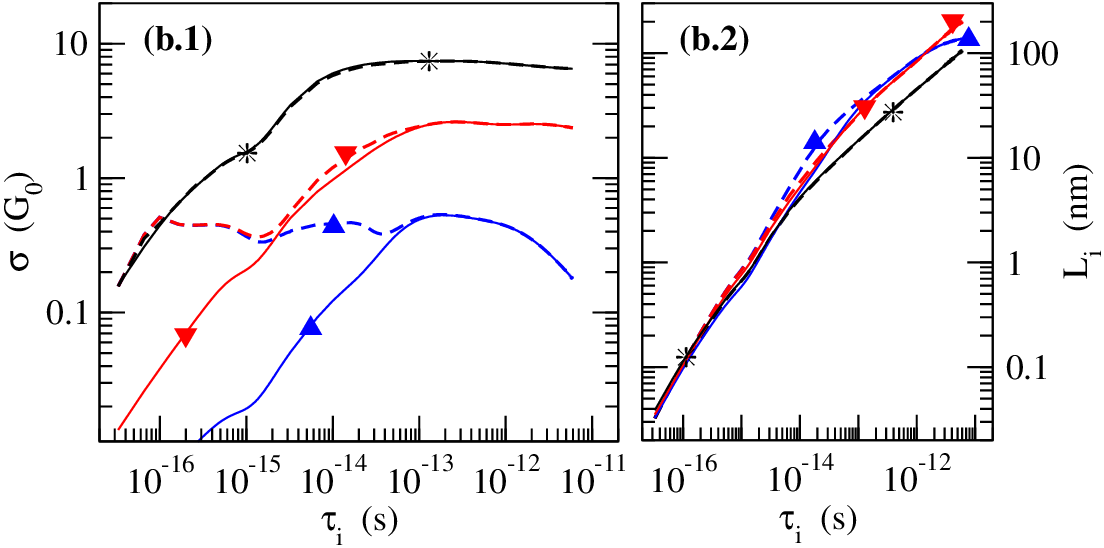}
\caption{ 
Conductivity $\sigma(E,\tau_{i})$ and inelastic mean-free path $L_{i}(E,\tau_{i})$
versus inelastic scattering time $\tau_{i}$. 
Same concentrations and energies as in figure \ref{figsup1}.
(a) Concentration $0.2\%$ of resonant adsorbates (monovacancies) for energies
 $E=0$ (triangle up),  $E=0.04$\,eV (triangle down) and  $E=0.8$\,eV (star).
(b) Concentration $1\%$ of non resonant adsorbates (divacancies) for energies 
  $E=0$ (triangle up), $E=0.1$\,eV (triangle down) and  $E=1.5$\,eV (star).
(Lines) without mixing on energy,
(dashed lines) with mixing on an energy range $\delta E = \hbar / \tau_i$,  i.e. $\eta = 1$ (see text).
$G_{0} = 2e^2/h$.
}
\label{f2}
\end{figure}

\begin{figure}
\includegraphics[clip,width=0.38\textwidth]{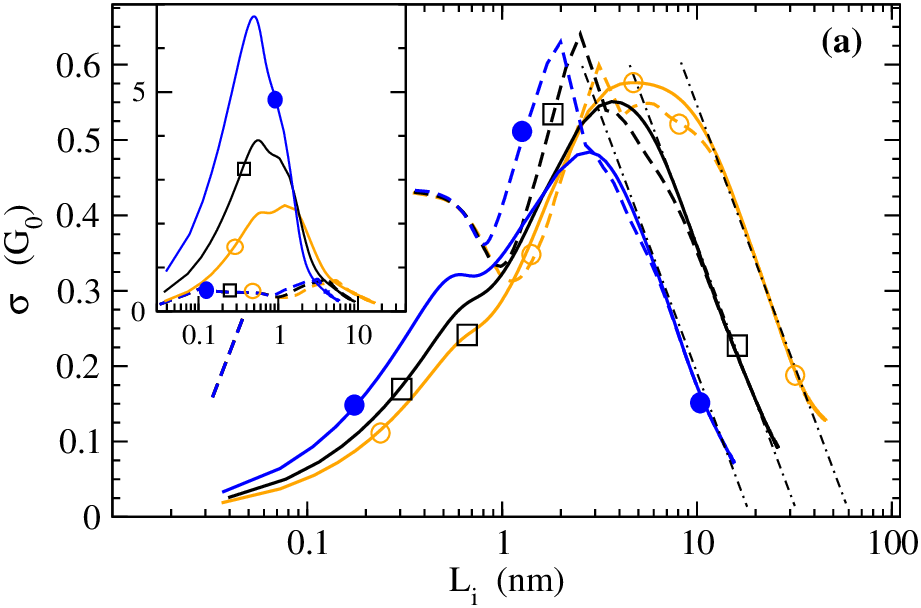}

\vskip .1cm
\includegraphics[clip,width=0.38\textwidth]{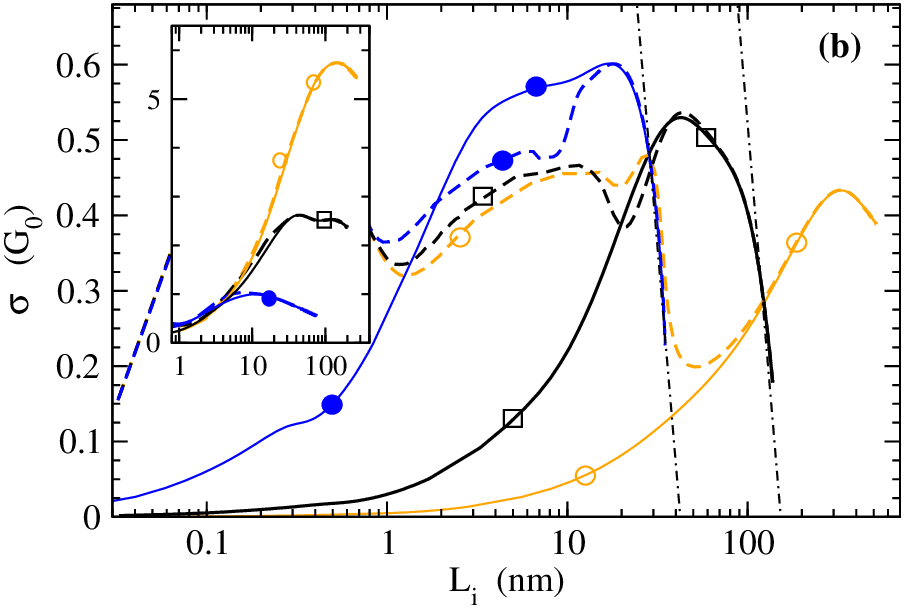}
\caption{Conductivity $\sigma$ in units of $G_{0} = 2e^2/h$ as a function of inelastic scattering length $L_{i}$
for the 3 concentrations of adsorbates (with the same symbols as in the main text figures 2(b) and 4(b)).
(a) resonant adsorbates (monovacancies) at energies  $E=0.03$\,eV.  
(b) non resonant adsorbates (divacancies) at  $E=0$\,eV.
(Continuous lines) with $\delta E = 0$, 
(dashed lines) with $\delta E = \hbar / \tau_{\rm i}$, i.e. $\eta = 1$ (see text).
The dot-dashed straight  lines show the slope $-\alpha$ for $L_{i} \gg L_{e}$ (equation (\ref{sigma})): 
(a) $\alpha=0.25$ and 
(b) $\alpha=0.75$.
The inset shows
(a) $\sigma(L_{i})$ at $E=0$ 
and (b) $\sigma(L_{i})$ at $E=0.1$\,eV. 
}
\label{figsup4}
\end{figure}

In the presence of inelastic scattering the long time propagation becomes diffusive as explained in the main text and in the second section  of this supplementary material. However there is another effect of the inelastic scattering. Indeed due to this scattering the eigenstates of the static Hamiltonian are no more eigenstates of the real system. 
Typically 
one expects that states in the energy range $E_{F} \pm \delta E /2$ with $\delta E = \eta  \hbar/\tau_{i}$ 
are mixing. 
$\eta = \tau_i / \tau_{im}$ is the ratio between the cutoff time $\tau_i$ of the weak localization (dephasing time)
and the inelastic scattering time $\tau_{im}$ corresponding to mixing of the DOS.
If the density of states and the diffusivity change quickly with energy this effect may be important. 
$\eta$ can depend on the system under consideration and we have at present no way to estimate it exactly.
Therefore we analyze this effect by computing the average of the density of states and of the conductivity  in the energy window $E_{F} \pm  \delta E /2$ with $\delta E = \eta  \hbar/\tau_{i}$  with $\eta$ varying from $1/100$ to 10, 
\beq
\left \langle n(E) \right \rangle_{\eta} = \frac{1}{ \delta E } \int_{E- \delta E / 2}^{E+ \delta E / 2} n(u) {\rm d}u \,,
\label{sigma3}
\eeq
\beq
\left \langle \sigma (E,\tau_{i}) \right \rangle_{\eta} = \frac{1}{ \delta E} \int_{E- \delta E /2 }^{E+ \delta E / 2} \sigma (u,\tau_{i}) {\rm d}u \, .
\label{sigma2}
\eeq
We find that this effect plays a minor role except for resonant adsorbates with strong inelastic scattering and very close to the Dirac energy where both the density of states and the microscopic conductivity are modified by the mixing of states (see figures \ref{figsup2} and \ref{figsup3}). As shown in figures \ref{f2} and \ref{figsup4} the long time regime which manifests the effect of Anderson localization is insensitive to the effect of mixing of states.

\end{document}